\documentclass{elsart}

\usepackage{latexsym}
\usepackage[dvips]{graphicx}

\usepackage{epsfig}
\usepackage{changebar}
\usepackage{amssymb}
\usepackage{amsmath}
\usepackage{amsbsy}
\usepackage{subeqnarray}
\usepackage{citesort}

\newcommand{\mc}{\multicolumn}
\newcommand{\lsim}{\mathrel{\mathop{\kern 0pt \rlap
  {\raise.2ex\hbox{$<$}}}
  \lower.9ex\hbox{\kern-.190em $\sim$}}}
\newcommand{\gsim}{\mathrel{\mathop{\kern 0pt \rlap
  {\raise.2ex\hbox{$>$}}}
  \lower.9ex\hbox{\kern-.190em $\sim$}}}

\begin{document}

\begin{frontmatter}

\title{Two-pion-exchange in the non-mesonic weak decay
of $\Lambda$-hypernuclei}

\author{C.~Chumillas$^1$, G.~Garbarino$^2$, A. Parre\~no$^1$ and A.
Ramos$^1$}

\address{$^1$Departament d'Estructura i Constituents de la Mat\`{e}ria and
\\ Institut de Ci\`encies del Cosmos,
\\Universitat de Barcelona, E-08028 Barcelona, Spain}

\address{$^2$Dipartimento di Fisica Teorica, Universit\`a di Torino and
INFN, \\ Sezione di Torino, I-10125 Torino, Italy}

\date{\today}

\begin{abstract}

The non-mesonic weak decay of $\Lambda$--hypernuclei is studied within a
one-meson-exchange potential supplemented by a chirally motivated
two-pion-exchange mechanism. The effects of final state interactions
on the outgoing nucleons are also taken into account. In view of the severe
discrepancies between theoretical expectations and experimental
data, particular attention is payed to the
asymmetry of the protons emitted by polarized hypernuclei.
The one-meson-exchange model describes the non-mesonic rates
and the neutron-to-proton ratio satisfactorily but predicts a too
large and negative asymmetry parameter. The 
uncorrelated and correlated two-pion mechanisms change the rates
moderately, thus maintaining the agreement with experiment. The
modification in the strength and sign of some decay amplitudes becomes
crucial and produces asymmetry parameters which lie well within
the experimental observations.
\end{abstract}

\begin{keyword}
Non-Mesonic Weak Decay \sep Decay Asymmetries of Polarized Hypernuclei
\sep $\Gamma_n/\Gamma_p$ ratio
\PACS 21.80.+a, 25.80.Pw, 13.30.Eg
%21.80.+a 	Hypernuclei
%13.30.Eg 	Hadronic decays
%13.75.Ev 	Hyperon-nucleon interactions
%13.88.+e 	Polarization in interactions and scattering
%24.70.+s 	Polarization phenomena in reactions
%25.40.-h 	Nucleon-induced reactions
%25.80.Pw 	Hyperon-induced reactions
\end{keyword}

\end{frontmatter}

%%%%%%%%%%%%%%%%%%%%%%%%%%%%%%%%%%%%%%%%%%%%%%%%%%%%%%%
\section{Introduction}
%%%%%%%%%%%%%%%%%%%%%%%%%%%%%%%%%%%%%%%%%%%%%%%%%%%%%%%
\label{intro}
Hypernuclei may be considered as a powerful ``laboratory" for
unique investigations of the baryon-baryon strangeness-changing
weak interactions. The field of non-mesonic weak decay
has indeed experienced a phase of renewed interest in the last few years.
%On the one hand, different theoretical \cite{Sa00,Os01,Pa02,It02,prl-prc,Bau06}
On the one hand, different theoretical \cite{Sa00,Os01,Pa97,Pa02,It02,prl-prc,Bau06}
and experimental \cite{Ok04,OutaVa,Ka06,Kim06}
indications have recently appeared in favour of a solution of the well-known and
long-standing puzzle on the ratio $\Gamma_n/\Gamma_p$ \cite{Ra98,Al02}
between the rates for the
$\Lambda n\to nn$ and $\Lambda p\to np$ non-mesonic weak decay processes.
Nowadays, values of this neutron-to-proton ratio around 0.3-0.4
for $s$- and $p$-shell hypernuclei are common to
both theoretical and experimental analyses.
An important role in this achievement has been played by a non-trivial
interpretation of data, which required analyses of two-nucleon induced
decays, $\Lambda NN\to nNN$, and accurate studies of nuclear
medium effects on the weak decay nucleons.
On the other hand, considerable concern is rooted in the persistence of
another open problem, of more recent origin,
which regards an asymmetry in the non-mesonic weak decay
of polarized hypernuclei and whose solution is expected to provide new
constraints for a deeper understanding of the dynamics of hypernuclear decay.
New phenomenological information is
in principle accessible since this asymmetry in the angular emission of protons
originates from the interference among parity-conserving (PC) and
parity-violating (PV) $\vec \Lambda p\to np$ transitions amplitudes \cite{Ba90,Ra92},
while the widely considered decay widths $\Gamma_n$ and $\Gamma_p$ are
the result of the incoherent sum of PC and PV amplitudes squared.
%dominated by
%the parity-conserving part of the interaction {\bf (Duda: siempre se dice eso
%del PC, y es verdad para OPE (70\% del NM es PC), pero los otros mesones pueden
%cambiar el \% de PC, invalidando lo que se dice!)}.

While inexplicable inconsistencies appeared between
the first asymmetry experiments of Refs.~\cite{Aj92,Aj00},
as discussed in Ref.~\cite{Al02}, very recent
and more accurate data \cite{OutaVa,Ma05,Ma06}
favour small \emph{observable} asymmetries, compatible with a vanishing value,
for both $s$- and $p$-shell hypernuclei. On the contrary, theoretical
%models based on one-meson-exchange potentials \cite{Sa00,Pa02,asyth05,Ba05}
models based on one-meson-exchange potentials \cite{Sa00,Pa97,Pa02,asyth05,Ba05}
(generally including the pseudoscalar and vector mesons
$\pi$, $\rho$, $K$, $K^*$, $\omega$ and $\eta$)
and/or direct quark mechanisms \cite{Sa00} predicted rather large and
negative \emph{intrinsic} asymmetry values (from $-0.7$ to $-0.5$
in the above quoted studies).
It must be noted that, on the contrary, the mentioned models
have been able to account fairly well for the total non-mesonic decay rates 
and for the
ratios $\Gamma_n/\Gamma_p$ measured for $s$- and $p$-shell
hypernuclei. Note also that, when discussing the comparison between theory
and experiment for the asymmetric non-mesonic decay, one should distinguish
(as we have done above) between \emph{intrinsic} and \emph{observable}
asymmetries \cite{asyth05}. Due
to nucleon final state interactions acting after the decay,
the former ($a_\Lambda$) is of theoretical relevance only, while
the real observable is the latter ($a^{\rm M}_\Lambda$).
A strong dependence of $a^{\rm M}_\Lambda$ on final state interactions
and proton detection threshold $T^{\rm th}_p$ has been obtained in Ref.~\cite{asyth05},
with values of $a^{\rm M}_\Lambda/a_\Lambda$ for standard experimental
conditions ($T^{\rm th}_p\simeq 40$ MeV) of about 0.7 for
$^{5}_\Lambda$He and 0.6 for $^{11}_\Lambda$B and $^{12}_\Lambda$C.

Recently, an effective field theory approach based on tree-level
pion- and kaon-exchange and leading-order contact interactions
has been applied
to hypernuclear decay \cite{Pa04}. The coefficients of the considered
four-fermion point interaction Lagrangian have been fitted
to reproduce the total and partial decay widths for $^{5}_\Lambda$He,
$^{11}_\Lambda$B and $^{12}_\Lambda$C, and the asymmetry parameter
for $^{5}_\Lambda$He. In this way,
a dominating central, spin-and isospin-independent
contact term has been predicted. Such term
turned out to be particularly important in order to reproduce a small
and positive value of the intrinsic asymmetry for $^{5}_\Lambda$He,
as indicated by the recent KEK experiments.
In order to improve the comparison with the observed decay asymmetries
in a calculation scheme based on a meson-exchange model,
this result could be interpreted dynamically as the need for the introduction
of a scalar-isoscalar meson-exchange.

Prompted by the work of Ref.~\cite{Pa04}, in Ref.~\cite{Sa05} a model based on
$(\pi+K)$-exchange and the direct quark mechanism has been supplemented with the
exchange of the scalar-isoscalar $\sigma$-meson.
Rather curiously, to date only a few works considered such a meson
as mediator of the non-mesonic decay, despite it could in principle be
relevant, being the chiral partner of the pion in QCD and
given that its mass is comparable with the exchanged momenta in the
$\Lambda N\to nN$ process
and with the mass of the kaon, a meson whose importance in accounting for the
%decay rates is, on the contrary, well established \cite{Sa00,Os01,Pa97}.
decay rates is, on the contrary, well established \cite{Sa00,Os01,Pa02}.
The strategy of Ref.~\cite{Sa05}
has been to determine the weak couplings of the $\sigma$ by fitting
decay data for $s$-shell hypernuclei.
The $\pi+K+\sigma$ $+$ direct quark model turned out to reproduce
the data for $\Gamma_{\rm NM}=\Gamma_n+\Gamma_p$, $\Gamma_n/\Gamma_p$
and $a^{\rm M}_\Lambda$ for $^5_\Lambda$He quite reasonably,
while the $\pi+K+\sigma$ model was unable to account for the experimental
value of $a^{\rm M}_\Lambda(^5_\Lambda {\rm He})$ \cite{foot1}.
Although the same model should also be tested for $p$-shell hypernuclei,
the results of Ref.~\cite{Sa05} clearly demonstrate the importance of the
$\sigma$-exchange in the non-mesonic decay.

A one-meson-exchange potential containing $\pi$, $\rho$, $K$, $K^*$, $\omega$,
$\eta$ and $\sigma$ has been applied more recently \cite{Ba06} to the
evaluation of $\Gamma_{\rm NM}$, $\Gamma_n/\Gamma_p$ and $a_\Lambda$
for $^5_\Lambda$He and $^{12}_\Lambda$C. The
unknown $\sigma$ couplings have been fixed to reproduce a subset
of decay observables [$\Gamma_{\rm NM}$ and $\Gamma_n/\Gamma_p$
for $^5_\Lambda$He], while the remaining observables have been predicted and
compared with data. The authors found that, despite the inclusion
of the $\sigma$ meson improved the overall agreement with experiment, the
asymmetry data for $^5_\Lambda$He could not be reproduced \cite{foot1}.

The contributions of
uncorrelated and correlated two-pion-exchange to the non-mesonic
weak decay has also been studied in Refs.~\cite{Os01,It02,Itproc,It04}.
Some preliminary papers \cite{Itproc} paved the way for a model that considered,
in addition to $\pi$ and $\omega$,
the exchange of two-pions
correlated in the scalar-isoscalar ($2\pi/\sigma$)
and vector-isovector ($2\pi/\rho$) channels \cite{It02}, 
with a phenomenological
treatment which is quite similar to the scheme used in the pioneering work of 
Ref.~\cite{Sh94}.
%in the first evaluation of correlated two-pion-exchange of Ref.~\cite{Sh94}.
The results of Ref.~\cite{It02} demonstrate how the correlated two-pion-exchange improves the
calculation of $\Gamma_n/\Gamma_p$ over the one-pion-exchange model.
After adding the exchange of the kaon, within a $\pi+K+\omega+2\pi/\sigma+2\pi/\rho$
model it was found \cite{It04} that the correlated two-pion-exchange also
entails some improvement in the evaluation of the asymmetry parameter.
In Ref.~\cite{Os01}, a meson-exchange potential including
pion, kaon, omega and uncorrelated plus correlated two-pions
($\pi+K+\omega+2\pi+2\pi/\sigma$) has been considered.
The correlated two-pion-exchange in the scalar-isoscalar
channel has been treated in terms of a chiral unitary approach
which has revealed to reproduce well $\pi \pi$ scattering data in the scalar sector
%, with a dynamical generation of the $\sigma$-meson resonance.
and in which the $\sigma$-meson appears as a dynamically generated resonance.
The only free parameter of the model is the momentum cutoff that regularizes
the loop integrals, which is fixed to about 1 GeV in order to
produce a $\sigma$ resonance at the observed mass of 450 MeV
and having a large width.
A sizable cancellation between $2\pi$- and $2\pi/\sigma$-exchange was
found for the relevant momenta ($\simeq 410$ MeV) in
the non-mesonic decay. Consequently, the total
two-pion-exchange contribution to the decay rates turned out to be moderate
but its effect on the asymmetry parameter was not evaluated.

%very very recently... [Itonaga at Hyp06] on $a_1$ meson...vanishing asymmetries
%for helium and carbon.....

Motivated by the findings of Refs.~\cite{Os01,Pa04,Sa05,It04,Ba06},
in the present paper we investigate the effects on the non-mesonic decay
observables for $s$- and $p$-shell hypernuclei
of uncorrelated two-pion-exchange and correlated two-pion-exchange in the
scalar-isoscalar channel.  We employ a finite nucleus approach and
pay special attention to the proton asymmetry.
The weak transition potentials for two-pion-exchange
are adopted from Refs.~\cite{Os01} and are added to the exchange
of the pseudoscalar and vector mesons $\pi$, $\rho$, $K$, $K^*$,
$\omega$ and $\eta$, with potentials taken from Ref.~\cite{Pa97}.
Correlated two-pion-exchange in the vector-isovector channel
is not considered here, since the one-meson-exchange
potential we use already includes the $\rho$-meson. We do not take into
account the two-nucleon induced decay mode, $\Lambda NN\to nNN$
\cite{Ra98,Al02,Al91}.
This channel can be safely neglected when evaluating the decay asymmetries
\cite{asyth05} even if its contribution to the total non-mesonic
decay rate is significant. When comparing with data for this rate,
the following predictions obtained in Refs.~\cite{prl-prc,Bau06,Al00PRC}
should be taken into account:
$\Gamma_2/(\Gamma_n+\Gamma_p)\simeq 0.20$ for $^5_\Lambda$He
and $\Gamma_2/(\Gamma_n+\Gamma_p)\simeq 0.25$ for $^{11}_\Lambda$B
and $^{12}_\Lambda$C.

With the novel meson-exchange model introduced in the present paper
it turns out to be possible to reproduce quite reasonably
experimental data for both the decay rates ($\Gamma_{NM}$ and
$\Gamma_n/\Gamma_p$) and the observable asymmetry parameter in $s$-
and $p$-shell hypernuclei. The introduction
of two-pion-exchange reveals to be of great importance in
this achievement.

The paper is organized as follows. The formalism employed
for the calculation of decay rates and asymmetries
is outlined in Section~\ref{formalism}. Numerical results for these observables
are presented and compared with data in Section~\ref{results}.
Finally, in Section~\ref{conclusion} we draw our conclusions.

%%%%%%%%%%%%%%%%%%%%%%%%%%%%%%%%%%
\section{Formalism}
%%%%%%%%%%%%%%%%%%%%%%%%%%%%%%%%
\label{formalism}

In this Section we briefly present the formalism adopted for the calculation
of the non-mesonic weak decay rates and proton asymmetries.

%%%%%%%%%%%%%%%%%%%%%%%%%%%%%%%%%%%%%%%%%%%
\subsection{Non-mesonic decay rates}
%%%%%%%%%%%%%%%%%%%%%%%%%%%%%%%%%%%%%%%%%%%
\label{NMWD}

The rate associated to a neutron (proton) stimulated decay can
be evaluated by the following average:
\begin{equation}
\label{gammanp}
\Gamma_{n(p)}=\frac{1}{2J+1} \, \sum_{M_J} \, \sigma_{n(p)}(J,M_J) \, {\rm ,}
\end{equation}
in terms of the intensities $\sigma_{n(p)}(J,M_J)$ of  neutrons (protons)
emitted along the quantization axis in the non-mesonic decay of
a hypernucleus with third component $M_J$ of the total spin $J$.

%(amplitudes in the proton elicity frame with the proton going out along
%the quantization axis)...such that

Following the approach used in Ref.~\cite{Pa97}, we apply
standard nuclear structure techniques that allow us to write these
intensities in terms of two-body amplitudes involving a $\Lambda N$ pair in the initial
state and a $NN$ pair in the final state. To do this, one needs to decouple the hyperon
(with spin and isospin quantum numbers $j_\Lambda, m_\Lambda, t_\Lambda=0, t_{3_\Lambda}=0$)
from the nucleon core ($J_C, M_C, T_I, T_{3_I}$) and the interacting
nucleon ($j_N, m_N, t_N=\frac{1}{2}, t_{3_N}$) from the residual system
($J_R, M_R, T_R, T_{3_R}$). Moreover, a sum over the quantum numbers
of the particles in the final state has to be performed. In terms of the
total and relative momenta of the two-nucleon final state,
$\vec P_T=\vec k_1+\vec k_2$ and $\vec k_r=(\vec k_1-\vec k_2)/2$, and 
working in a coupled two-body spin-isospin basis, one has:
\begin{eqnarray}
\label{sigmaN}
\sigma_N(J,M_J)
&=& \int \frac{d^3 P_T}{(2\pi)^3} \, \int \frac{d^3 k_r}{(2\pi)
^3}\,
(2\pi) \, \delta(M_H-E_R-E_1-E_2) \nonumber \\
&\times& \sum_{S M_S} \sum_{J_R M_R} \sum_{T_R T_{3_R}}
\mid \langle T_R T_{3_R}, \frac{1}{2} t_{3_N}
\mid T_I T_{3_I} \rangle \mid^2  \nonumber \\
&\times& \biggl| \, \sum_{T T_3}
\langle T T_3 \mid \frac{1}{2} -\frac{1}{2}, \frac{1}{2} t_{3_N} \rangle
\sum_{m_\Lambda M_C} \langle j_\Lambda m_\Lambda, J_C M_C \mid
J M_J \rangle \label{eq:intensities}  \\
&\times& \sum_{j_N} S^{1/2}( J_C \, T_I \, ; J_R \, T_R \, , j_N \,t_{3_N} )
\sum_{M_R m_N} \langle J_R M_R, j_N m_N \mid J_C M_C\rangle \nonumber \\
&\times& \sum_{m_{l_N} m_{s_N}} \langle j_N m_N \mid l_N m_{l_N},
\frac{1}{2} m_{s_N} \rangle  \,
\sum_{m_{l_\Lambda} m_{s_\Lambda}} \,\,
\langle j_\Lambda m_\Lambda \mid l_\Lambda m_{l_\Lambda}, \frac{1}{2}
m_{s_\Lambda} \rangle \nonumber \\
&\times& \sum_{S_0 M_{S_0}} \langle S_0 M_{S_0} \mid \frac{1}{2} m_{s_\Lambda}, \frac{1}{2}
m_{s_N} \rangle \,
\sum_{T_0 T_{3_0}} \langle T_0 T_{3_0} \mid \frac{1}{2}\, -\frac{1}{2},
\frac{1}{2} t_{3_N} \rangle
\nonumber \\
&\times& \frac{1-(-1)^{(L+S+T)}}{\sqrt{2}} \nonumber \\ 
&\times& t_{\Lambda N \to nN}
(S,M_S,T,T_3,S_0,M_{S_0},T_0,T_{3_0},l_\Lambda,l_N,{\vec P}_T,
{\vec k}_r) \biggl|^2 \, {\rm .} \nonumber
\end{eqnarray}
In Eq.~(\ref{eq:intensities}), $M_H$ stands for the mass of the initial hypernucleus,
which is assumed to be at rest, $E_1, E_2$ and $E_R$ are the asymptotic total energies of the
two nucleons and the residual nucleus in the final state and
$S^{1/2}( J_C \, T_I \, ; J_R \, T_R \, , j_N \,t_{3_N})$ is a
nucleon pick-up spectroscopic amplitude. The index $N=n$ or $p$ determines if the
decay is induced by a neutron or a proton, with
$t_{3n}=-1/2$ and $t_{3p}=1/2$ correspondingly. The elementary amplitude
$t_{\Lambda N \to nN}$ accounts for the transition
from an initial $\Lambda N$ state with spin (isospin) $S_0$ $(T_0)$
to a final antisymmetric $nN$ state with spin (isospin) $S$ $(T)$.
It can be written in terms of other elementary amplitudes which depend on center-of-mass
($N_R,L_R$) and relative ($N_r,L_r$) principal and orbital angular momentum
quantum numbers of the $\Lambda N$ and $nN$ systems:
\begin{eqnarray}
t_{\Lambda N \to nN} =\sum_{N_r L_r N_R L_R} X(N_r L_r, N_R L_R, l_\Lambda l_N)
\, t_{\Lambda N \to nN}^{N_r L_r\, N_R L_R} \ ,
\label{moshinski}
\end{eqnarray}
where the dependence on the spin and isospin quantum numbers
has to be understood. In Eq.~(\ref{moshinski}), the coefficients
$X(N_r L_r, N_R L_R, l_\Lambda l_N)$ are the well known Moshinsky brackets,
 while:
\begin{eqnarray}
t_{\Lambda N \to nN}^{N_r L_r\, N_R L_R} &=&
\frac{1}{\sqrt{2}}\int d^3 R \int d^3 r \, {\rm e}^{- i {\vec P}_T\cdot {\vec R}}
\Psi^*_{{\vec k}_r}\, ({\vec r}\,) \chi^{\dagger\, S}_{M_S}
\chi^{\dagger\, T}_{T_3} \nonumber \\
&\times& V_{\sigma,  \tau}({\vec r}\,) \, \Phi^{\rm CM}_{N_R L_R}\left(\frac{{\vec
R}}{b/\sqrt{2}}\right) \Phi^{\rm rel}_{N_r L_r}\left(\frac{{\vec r}}{\sqrt{
2}b}\right)
\chi^{S_0}_{M_{S_0}} \chi^{T_0}_{T_{3_0}} \, {\rm ,}
\label{erre-dep}
\end{eqnarray}
with $V_{\sigma,  \tau}({\vec r}\,)$ the weak transition potential depending on the relative
coordinate between the interacting $\Lambda$ and nucleon, $r$, and their spin,
$\sigma$, and
isospin, $\tau$, variables. Moreover, $\Phi^{\rm rel}_{N_r L_r}({\vec r}/(\sqrt{2}b))$ and
$\Phi^{\rm CM}_{N_R L_R} ({\vec R}/(b/\sqrt{2}))$ are the
relative and center--of--mass harmonic oscillator wave functions
describing the $\Lambda N$ system, while
$\Psi_{{\vec k}_r}\, ({\vec r}\,)$ and ${\rm e}^{i {\vec P}_T\cdot {\vec R}}$
are the relative and center-of-mass wave functions of the $NN$ final state.

When extracting information on the elementary weak two-body interaction taking
place in the medium, it is vital to account for the strong interaction between
the hadrons in the initial and final states. The wave function $\Psi_{{\vec
k}_r}\, ({\vec r}\,)$, describing the relative motion of the two nucleons under
the influence of a suitable $NN$ interaction, is obtained from the
Lippmann-Schwinger  equation. For the initial $\Lambda N$ system we start from
a mean field approach where the $\Lambda$ and nucleon single particle wave
functions are obtained from harmonic oscillator potentials. Their corresponding
oscillator parameters have been adjusted to reproduce the $\Lambda$ separation
energy in the hypernucleus under consideration and the charge form factor for
the corresponding nuclear core. To obtain the correlated $\Lambda N$ wave function one should
solve a similar equation as in the $NN$ case,  but with the Pauli operator
acting on the propagation of the intermediate states properly  incorporated
($G-$matrix equation). A simpler approach, which has been tested in the weak
decay  of $^5_\Lambda$He \cite{Pa97}, consists in fitting microscopic
$G-$matrix calculations with a phenomenological spin-independent correlation
function that multiplies the harmonic
oscillator $\Lambda N$ wave function to obtain the correlated one. For the
baryon-baryon strong interactions we take the Nijmegen soft-core model, version
NSC97f~\cite{nsc97f}, which has been used with success in hypernuclear structure
calculations as
well as in the decay  of hypernuclei.

\begin{figure}[h]
\begin{center}
\includegraphics[width=8cm]{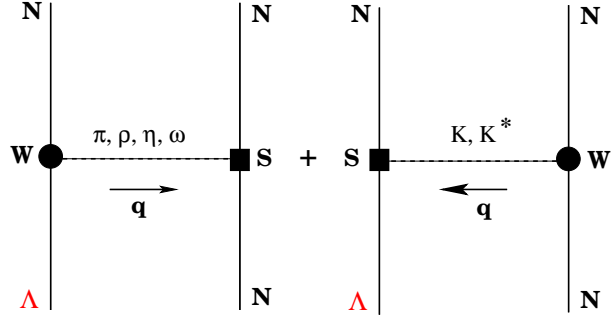}
\caption{Feynman diagrams corresponding to the weak $|\Delta S|=1$ $\Lambda N \to NN$
transition amplitudes mediated by the exchange of the pseudoscalar $\pi, \eta, K$ mesons
and the vector $\rho, \omega, K^*$ mesons. The circle and the square stand for the weak and
strong vertices, respectively.}
\label{ome}
\end{center}
\end{figure}

Within our One-Meson-Exchange (OME) model, the weak transition potential is built from the
exchange of virtual mesons belonging to the ground state pseudoscalar and vector octets,
$\pi, \eta, K, \rho, \omega, K^*$, as depicted in Fig.~\ref{ome}. For the sake of simplicity,
we will not derive here the expression for the transition potential starting from the weak
and strong vertices entering each Feynman amplitude; details on this calculation can be found
in Ref.~\cite{Pa97}. Here we only quote the final result for the (non-relativistic)
potential, which is:
\begin{eqnarray}
V_{\sigma,  \tau} ({\vec r}\,) &=&
\sum_i \sum_\alpha V_\alpha^{(i)} ({\vec r}\,) =
\sum_i \sum_\alpha V_\alpha^{(i)} (r) \hat{O}_\alpha (\vec{\sigma}, \hat{r}) \,
\hat{I}_\alpha^{(i)} \nonumber \\
&=& \sum_i [V_C^{(i)} (r) \hat{I}_C^{(i)} + V_{SS}^{(i)} (r) \,
{\vec \sigma}_1 \cdot {\vec \sigma}_2 \, \hat{I}_{SS}^{(i)}
+ V_T^{(i)}(r) \, S_{12}(\hat{r}) \, \hat{I}_T^{(i)} \nonumber \\
&+&
(n^i {\vec \sigma}_2 \cdot {\vec r} + (1-n^i)[{\vec \sigma}_1 \times {\vec \sigma}_2]
\cdot {\vec r}\,) V_{PV}^{(i)}(r) \hat{I}_{PV}^{(i)}]\, {\rm ,}
\end{eqnarray}
where $S_{12}(\hat{r}) = 3 \,{\vec \sigma}_1 \cdot {\hat r}\, {\vec \sigma}_2 \cdot {\hat r} -
{\vec \sigma}_1 \cdot {\vec \sigma}_2$ is the tensor operator and
$n^i=1(0)$ for pseudoscalar (vector) mesons. The index $i$ runs over the different exchanged
mesons and the index $\alpha$ over the different transition channels, central spin-independent,
central spin-dependent, tensor and parity violating.
Again, to avoid an excess of information, we refer to Ref.~\cite{Pa97} for the explicit
form of $V_\alpha^{(i)} (r)$, for the isospin factors $\hat{I}_\alpha^{(i)}$, as well as for the numerical values of the coupling constants.

\begin{figure}[h]
\begin{center}
\includegraphics[width=8cm]{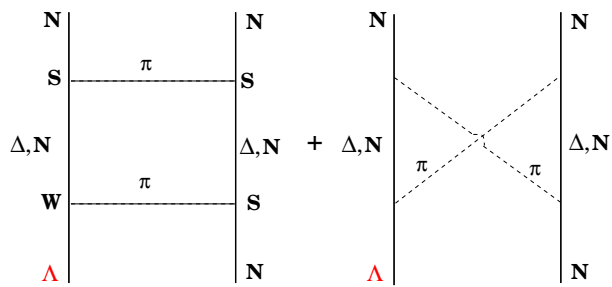}
\caption{Uncorrelated two-pion exchange diagrams, direct and crossed, for the weak
$|\Delta S|=1$ $\Lambda N \to NN$ transition amplitude. Only $\Delta N$ and
$\Delta \Delta$ intermediate states are allowed to avoid double counting (see explanation in the
text).}
\label{2pion_unc}
\end{center}
\end{figure}

\begin{figure}[h]
\begin{center}
\includegraphics[width=5cm]{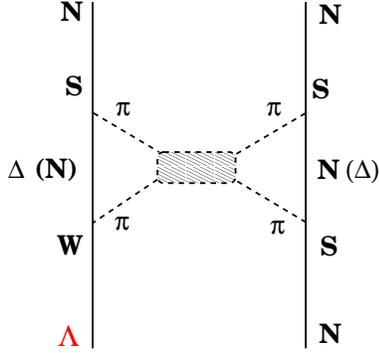}
\caption{Correlated two-pion exchange diagram for the weak $|\Delta S|=1$
$\Lambda N \to NN$ transition amplitude.}
\label{2pion_cor}
\end{center}
\end{figure}

In the present work we complement this OME potential with the
contributions of uncorrelated ($2\pi$) and correlated
($2\pi/\sigma$) two-pion-exchange taken from Ref.~\cite{Os01}. In
that work, a chiral unitary model has been used to account for the
correlated two-pion-exchange in the scalar-isoscalar channel. This
scheme was built originally for the nucleon-nucleon interaction,
leading to a $2\pi/\sigma$-exchange potential with a moderate
attraction at $r\gsim 0.9$~fm and a repulsion at shorter distances
\cite{Os00}, in contrast with the attraction at all distances of
the standard phenomenological $\sigma$-meson exchange. Once the
uncorrelated and correlated two-pion-exchange are added together,
an attractive nucleon-nucleon potential is obtained for all
distances. Applying an appropriate conversion factor that replaces
a strong $\pi NN$ vertex with the weak $\pi \Lambda N$ one, the
potential was implemented in the study of the weak decay of
hypernuclei \cite{Os01} .
%In order to restore the behavior of realistic
%nucleon-nucleon potentials, which present a moderate
%attraction only at intermediate distances,
%the authors of Ref.~\cite{Os01} introduced the exchange of the $\omega$-meson
%to produce the required repulsion.
The relevant diagrams for uncorrelated and correlated two-pion-exchange with
intermediate $N$ and $\Delta$ states built in Ref.~\cite{Os01} are depicted in
Figs.~\ref{2pion_unc}
and ~\ref{2pion_cor}, respectively. Two-nucleon intermediate states are not
considered in the uncorrelated (direct and crossed) diagrams in order to
avoid double counting when including the $NN$ strong correlations via the solution of
a Lippmann-Schwinger equation, which includes the contribution of iterated
one-pion-exchange interactions. In was found that, in the range of momenta relevant for
the non-mesonic weak decay, the results from the
uncorrelated two-pion diagrams with  intermediate $\Delta N$ and $\Delta\Delta$ 
states are largely dominated by the isoscalar piece, which is the only one retained in their
final results. As for the $2\pi/\sigma$ correlated contribution,
the box in Fig.~\ref{2pion_cor} contains the pion-pion scattering
$t$-matrix summed up to all orders in the unitary approach.
Diagrams containing intermediate $\Sigma$ and $\Sigma^*$ baryons are not
included since their individual contributions approximately cancel each other
when these baryons are considered together \cite{Os01}.
We also note that the complete scalar-isoscalar two-pion-exchange potential
given in Ref.~\cite{Os01} is of pure parity-conserving nature. The reason is
that the parity-violating contribution is strongly reduced by the lack of direct
coupling of the $\Lambda$ to $\Delta$ intermediate states.
The results of Ref.~\cite{Os01} show a large cancellation
between correlated $2\pi/\sigma$ and uncorrelated $2\pi$-exchange
at momentum values which are relevant for
the non-mesonic decay. Consequently, the total
two-pion-exchange contribution to the decay rate turns out to be small.

Calculations performed with the finite nucleus approach based on
Eqs.~(\ref{gammanp}) and (\ref{sigmaN}) and adopting the OME
potential previously mentioned reproduced quite well the
$\Gamma_n/\Gamma_p$ values determined from data on coincidence
nucleon spectra \cite{prl-prc}. However, as all the other OME
models employed to date, it failed in accounting for the recent
asymmetry data \cite{asyth05}. In the present paper we will see
that the implementation of two-pion-exchange contributions
modifies the decay widths moderately, as in Ref.~\cite{Os01}, but
has a tremendous influence on the decay asymmetries, bringing them to
values that are in perfect agreement with the recent experimental
data.

%%%%%%%%%%%%%%%%%%%%%%%%%%%%%%%%%%
\subsection{Asymmetry parameters}
%%%%%%%%%%%%%%%%%%%%%%%%%%%%%%%%%%
\label{Asy}

The study of polarized hypernuclei provides us with
new and complementary phenomenological insights on the
weak hyperon-nucleon interaction.
Indeed, while investigations of the decay rates basically
serve to clarify the isospin structure of the non-mesonic
transitions, asymmetry studies allow us to
%%From such investigations one can
extract significant information on the strength and the relative
phases of the different decay amplitudes.
%Indeed, while investigations of the decay rates basically
%serve to clarify the isospin structure of the non-mesonic
%transitions, {\bf especially for the dominating PC part ?,
%the decay asymmetries, depending on interferences between
%PC and PV transition amplitudes,
%can be more easily used to put constraints on the PV
%amplitudes.}

Using the reaction $n(\pi^+,K^+)\Lambda$ on a $^{12}$C  ($^6$Li) target at
KEK \cite{Aj92,Ma05}  (\cite{Aj00,Ma06}), 
with a pion momentum of $\sim 1.05$ GeV and small
kaon emission angles, $2^\circ \le \theta_K \le 15^\circ$,
$^{12}_\Lambda$C and $^{11}_\Lambda$B  ($^5_\Lambda$He) hypernuclei 
have been produced with large spin-polarization, aligned preferentially
along the axis normal to the reaction plane.
%Under these conditions, one can evaluate the intensities of neutrons and protons
%$\sigma_{n(p)}(J,M_J)$ of Eq.(\ref{sigmaN}) emitted along the quantization axis
%in the non-mesonic decay of a hypernucleus with third component $M_J$ of
%the total spin $J$.

%\begin{figure}[h]
%\includegraphics[width=10cm]{polar_reaction.eps}
%\caption{
%Schematic illustration of the $n(\pi^+,K^+)\Lambda$ reaction on $^{12}$C performed
%at KEK. The typical kinematical conditions of such reaction, with a pion
%momentum of $\sim 1.05$ GeV and small kaon angles with respect to the
%reaction plane $(2^\circ - 15^\circ)$,
%produces $^{12}_\Lambda$C hypernuclei with large spin-polarization aligned
%preferentially along the axis normal to the reaction plane. The PV asymmetry
%is obtained by looking at the angular distribution of the weak decay protons.}
%\label{polar}
%\end{figure}

The expression for the intrinsic asymmetry
parameter derived below follows the pioneering work developed in Ref.~\cite{Ra92}.
Neglecting for the moment nucleon final state interactions, the
intensity of protons from $\vec \Lambda p\to np$ decays emitted along
a direction forming an angle $\theta$ with respect to the
hypernuclear spin-polarization axis
%, as depicted in Fig.~\ref{polar},
is:
\begin{equation}
\label{int-w}
I(\theta,J)=I_0(J)\left[1+{\mathcal{A}}(\theta, J)\right]~,
\end{equation}
with
\begin{equation}
\label{int-w2}
{\mathcal{A}} (\theta,J) = P_y(J) A_y(J) \cos \theta~.
%= p_\Lambda(J)\, a_\Lambda\, \cos \theta~.
\end{equation}
In the above, $J$ is the hypernuclear total spin, $I_0$ the
isotropic intensity for an unpolarized hypernucleus:
\begin{equation}
\label{izero}
I_0(J)=\frac{1}{2J+1}\sum_{M_J}\sigma_p(J,M_J)\equiv \Gamma_p~,
\end{equation}
$P_y$ the hypernuclear polarization, which depends on the kinematics and
dynamics of the hypernuclear production reaction, and $A_y$ the hypernuclear
asymmetry parameter:
\begin{equation}
\label{ay}
A_y(J)=\frac{3}{J+1}\frac
{\displaystyle \sum_{M_J} M_J\, \sigma_p(J,M_J)}
{\displaystyle \sum_{M_J}\sigma_p(J,M_J)}~.
\end{equation}
The partial decay rates $\sigma_p(J,M)$ entering Eqs.~(\ref{izero})
and (\ref{ay}) are defined by Eq.~(\ref{eq:intensities}).
The shell model weak-coupling scheme, in which the $1s_{1/2}$ $\Lambda$
is assumed to be coupled to the nuclear core ground state with total spin $J_C$,
allows rewriting the asymmetry $\mathcal{A}$ in terms of the polarization of the
hyperon spin, $p_\Lambda$, together with the corresponding \emph{intrinsic} $\Lambda$
asymmetry parameter, $a_\Lambda$:
%\begin{eqnarray}
%\label{a-lambda}
%a_{\Lambda} &=&
%\begin{cases}
%-\displaystyle\frac{J+1}{J} A_y(J) & \text{if} \, \, \, J=J_C-\frac{1}{2} \\
% A_y(J) & \text{if} \, \, \, J=J_C +\frac{1}{2}
%\end{cases} \\
%& & \\
%\label{p-lambda}
%p_{\Lambda}(J) &=&
%\begin{cases}
%-\displaystyle\frac{J}{J+1} P_y(J) & \text{if}\, \, J=J_C-\frac{1}{2}  \\
%P_y(J) & \text{if}\, \, J=J_C+\frac{1}{2}
%\end{cases}
%\end{eqnarray}
\begin{equation}
p_\Lambda(J) =\left\{ \begin{array}{c c l}
-\displaystyle\frac{J}{J+1} P_y(J) &\phantom{AAAA}& {\rm if } \ \
J=J_C-\frac{1}{2} \\
 P_y(J) &\phantom{AAA}& {\rm if } \ \ J=J_C+\frac{1}{2} \end{array}
\right.           \ ,
\label{eq:polar}
\end{equation}
\begin{equation}
a_\Lambda =\left\{ \begin{array}{c c l}
-\displaystyle\frac{J+1}{J} A_y(J) &\phantom{AAAA}& {\rm if } \ \
J=J_C-\frac{1}{2} \\
 A_y(J) &\phantom{AAA}& {\rm if } \ \ J=J_C+\frac{1}{2} \end{array}
\right.           \ .
\label{eq:asym}
\end{equation}
In this way, Eq.~(\ref{int-w2}) becomes:
\begin{equation}
{\mathcal{A}} (\theta,J) = p_\Lambda(J)\, a_\Lambda\, \cos \theta~,
\end{equation}
and $a_\Lambda$ can be interpreted as being
an intrinsic attribute of the elementary $\vec \Lambda p\to np$ process, in
which case it should be practically independent of the decaying hypernucleus.
Several calculations \cite{Pa97,Pa02,Ra98,Al02,Ra92,asyth05,Ba05,Pa04,Ba06}
have indeed demonstrated that this asymmetry
shows only a moderate dependence on the hypernuclear structure.
%of the hypernucleus in which the non-mesonic decay takes place,
%thus revealing the limited validity of the weak-coupling approximation.
%One should also note that experimentally the directly measurable
%quantity is the product $P_y(J) A_y(J)$...

Nucleon final state interactions strongly modify the weak decay intensity
of Eqs.~(\ref{int-w}) and (\ref{int-w2}) and the
experimentally accessible quantity is an observable proton intensity
of the form \cite{asyth05}:
\begin{equation}
\label{int-exp}
I^{\rm M}(\theta,J)=I^{\rm M}_0(J)
[1+p_\Lambda(J)\, a^{\rm M}_\Lambda(J) \cos \theta]~.
\end{equation}
The corresponding \emph{observable} asymmetry is thus obtained from the measured
or calculated intensity as:
\begin{equation}
\label{a-exp}
a^{\rm M}_\Lambda (J)=\frac{1}{p_\Lambda(J)}\,
\frac{I^{\rm M}(0^{\circ},J)-I^{\rm M}(180^{\circ},J)}
{I^{\rm M}(0^{\circ},J)+I^{\rm M}(180^{\circ},J)}~,
\end{equation}
and in general depends on the considered hypernucleus and on
experimental conditions such as the adopted proton detection threshold.
From Eq.~(\ref{a-exp}) it is evident that to determine experimentally
$a^{\rm M}_\Lambda$, a measurement of
%the hypernuclear polarization $P_y$ [see Eq.~(\ref{p-lambda})] is required.
the hypernuclear polarization $P_y$ [see Eq.~(\ref{eq:polar})] is required.
Such a measurement has been possible for $^5_\Lambda$He \cite{Aj00},
but only theoretical evaluations of $P_y$ are available
for  $p$-shell hypernuclei \cite{Aj92,Ma05,Ma06}.

The relation between intrinsic and observable asymmetries
has been investigated for the first time in Ref.~\cite{asyth05}, where
the Monte Carlo intranuclear cascade model of
Ref.~\cite{Ra97} has been used to account for nucleon final state interactions.
%There, hypernuclear decay was computed in infinite nuclear matter, and the
%finite nucleus result
%was obtained through a Local Density Approximation.
%The approach includes also the $2N$-induced mechanism as dominated by the
%absorption on a $np-$correlated pair at the weak $\Lambda n \pi$ vertex. Its contribution
%was estimated within the polarization propagator method in LDA~\cite{ROS94,APGR00} to be
%around the $20\%$ for $s-$shell nuclei and $25\%$ for $p-$shell ones.
%The intranuclear cascade calculation will follow the fate of the 3 nucleons emitted
%afterwards.
%%%When applied to the present investigation,
This code is based on the following basic ingredients through which
the kinematics of the emitted nucleons is generated:
i) first, according to the values of $\Gamma_n$ and $\Gamma_p$
predicted by the adopted meson-exchange model,
a random number generator decides if the decay is neutron or proton-induced,
thus determining the charges of the weak decay nucleons;
ii) based on the relevant density and momentum probability distributions,
the same random number generator selects the positions and momenta
of these primary nucleons;
%and the properly scaled value of $\Gamma_{np}$~\footnote{The
%calculation mixes two different formalisms, the $1N-$induced OME formalism in finite nuclei
%of Ref.~\cite{Pa97}, and the Polarization Propagator
%Method in LDA to determine the $2N-$induced channel.
%This implies that the obtained distributions of the weak decay nucleons and the
%$\Gamma_{\rm 2N}$ value have been properly normalized to keep the
%$\displaystyle\frac{\Gamma_{\rm 2N}}{\Gamma_{\rm 1N}}$ unchanged,
%$\displaystyle\frac{\Gamma_{\rm 2N}}{\Gamma_{\rm 1N}} \equiv
%\left( \displaystyle\frac{\Gamma_{\rm 2N}}{\Gamma_{\rm 1N}} \right)^{\rm LDA} = 0.20$ for
%$^5_\Lambda$He and $0.25$ for $^{12}_\Lambda$C.}.
%After the primary nucleons are generated, they move under a local potential.
%$V_N (R) = - \displaystyle\frac{k_{F_N}^2 (R)}{2 m_N}$, with $k_{F_N}(R)$ the local Fermi
%momentum.
iii) next, they propagate under the influence of a local potential
and are allowed to collide with the nucleons of the medium, thus
producing, among other effects, the emission of secondary nucleons.
Each Monte Carlo event generates a certain number of nucleons
which leave the nucleus with definite momenta. With enough statistics, 
one can then build up
single nucleon spectra, coincidence spectra or up-down proton asymmetries 
that are direcly comparable with the expertimental observations.

%%%%%%%%%%%%%%%%%%%%%%%%%%%%%%%%%%
\section{Results}
%%%%%%%%%%%%%%%%%%%%%%%%%%%%%%%%
\label{results}

\begin{table}
\begin{center}
\caption{The non-mesonic weak decay rates (in units of the free $\Lambda$ decay width)
and intrinsic asymmetry parameters predicted for $^5_\Lambda$He, $^{11}_\Lambda$B
and $^{12}_\Lambda$C are compared with recent data. See text for details.}
\label{nmdw}
\resizebox*{\textwidth}{!}{
\begin{tabular}{l c c c c c}
\mc {1}{c}{} &
\mc {1}{c}{} &
\mc {1}{c}{} &
\mc {1}{c}{$^5_\Lambda$He} &
\mc {1}{c}{} &
\mc {1}{c}{} \\
%\mc {1}{c}{} &
%\mc {1}{c}{$\Gamma_n$} &
%\mc {1}{c}{$\Gamma_p$} &
%\mc {1}{c}{$\Gamma_{NM}=\Gamma_n+\Gamma_p$} &
%\mc {1}{c}{$\Gamma_n/\Gamma_p$} &
%\mc {1}{c}{$a_\Lambda$} \\ \hline
Model & ~$\Gamma_n$~~~ & ~~~$\Gamma_p$~ &
$\Gamma_{NM}=\Gamma_n+\Gamma_p$ & $\Gamma_n/\Gamma_p$ &
$a_\Lambda$\\ \hline
OME                   & 0.122 & 0.257 & 0.379 & 0.474 & $-0.590$\\
\hline
$\pi$                          & 0.040 & 0.420 & 0.460 & 0.095 & $-0.231$\\
$\pi+K$                        & 0.097 & 0.189 & 0.286 & 0.510 & $-0.544$\\
%$\pi+K+2\pi/\sigma$            & 0.104 & 0.206 & 0.310 & 0.501 & $-0.630$ \\
$\pi+K+2\pi$                   & 0.121 & 0.329 & 0.450 & 0.368 & $+0.181$ \\
$\pi+K+2\pi+2\pi/\sigma$       & 0.111 & 0.285 & 0.396 & 0.390 & $+0.114$\\
${\rm OME}+2\pi+2\pi/\sigma$   & 0.114 & 0.275 & 0.388 & 0.415 & $+0.041$\\
\hline
KEK-E462 \protect\cite{OutaVa,Ka06,Ma05} &  &  & $0.424\pm0.024$
& $0.45\pm 0.11\pm 0.03$ & $0.11\pm 0.08\pm 0.04$ \\
KEK-E462 \protect\cite{Ma06} &  &  & & & $0.07\pm 0.08^{+0.08}_{-0.00}$ \\
KEK-E462 \protect\cite{Ka06} &  &  & & $0.39\pm 0.11$ ($1N$)& \\
(analysis of Ref.~\protect\cite{prl-prc}) &  &  & & $0.26\pm 0.11$ ($1N+2N$) & \\ \hline\hline
 & & & $^{11}_\Lambda$B & & \\
Model & $\Gamma_n$ & $\Gamma_p$ & $\Gamma_{NM}=\Gamma_n+\Gamma_p$
& $\Gamma_n/\Gamma_p$ & $a_\Lambda$\\ \hline
OME            & 0.179 & 0.408 & 0.587 & 0.439 & $-0.809$\\
\hline
$\pi$                         & 0.067 & 0.619 & 0.685 & 0.108 & $-0.353$\\
$\pi+K$                       & 0.137 & 0.308 & 0.445 & 0.447 & $-0.773$\\
%$\pi+K+2\pi/\sigma$           & 0.150 & 0.335 & 0.485 & 0.447 & $-0.838$\\
$\pi+K+2\pi$                  & 0.199 & 0.480 & 0.678 & 0.414 & $+0.025$ \\
$\pi+K+2\pi+ 2\pi\sigma$      & 0.187 & 0.427 & 0.613 & 0.438 & $-0.074$\\
${\rm OME}+2\pi+2\pi/\sigma$  & 0.202 & 0.425 & 0.627 & 0.474 & $-0.181$\\
 \hline
KEK-E508 \protect\cite{OutaVa,Ma05} &  &  &  &  & $-0.20\pm 0.26 \pm 0.04$\\
KEK-E508 \protect\cite{Ma06} &  &  & & & $-0.16\pm 0.28^{+0.18}_{-0.00}$ \\
KEK-E307 \protect\cite{Sato05} & & & $0.861\pm 0.063\pm 0.073$ & & \\ \hline\hline
 & & & $^{12}_\Lambda$C & & \\
Model & $\Gamma_n$ & $\Gamma_p$ & $\Gamma_{NM}=\Gamma_n+\Gamma_p$
& $\Gamma_n/\Gamma_p$ & $a_\Lambda$\\ \hline
OME         & 0.175 & 0.491 & 0.667 & 0.357 & $-0.698$\\
\hline
$\pi$                          & 0.066 & 0.751 & 0.817 & 0.088 & $-0.350$\\
$\pi+K$                        & 0.134 & 0.371 & 0.505 & 0.363 & $-0.643$\\
%$\pi+K+2\pi/\sigma$            & 0.147 & 0.402 & 0.549 & 0.365 & $-0.722$ \\
$\pi+K+2\pi$                   & 0.195 & 0.581 & 0.776 & 0.335 & $+0.007$ \\
$\pi+K+2\pi+2\pi/\sigma$       & 0.182 & 0.521 & 0.703 & 0.349 & $-0.093$\\
${\rm OME}+2\pi+2\pi/\sigma$   & 0.194 & 0.529 & 0.722 & 0.366 & $-0.207$\\
\hline
KEK-E508 \protect\cite{OutaVa,Kim06,Ma05} &  &  & $0.940\pm 0.035$
& $0.51\pm 0.13\pm 0.05$ & $-0.20\pm 0.26\pm 0.04$ \\
KEK-E508 \protect\cite{Ma06} &  &  & & & $-0.16\pm 0.28^{+0.18}_{-0.00}$ \\
KEK-E508 \protect\cite{Kim06} &  &  & & $0.38\pm 0.14$ ($1N$) & \\
(analysis of Ref.~\protect\cite{prl-prc}) &  &  & & $0.29\pm 0.14$ ($1N+2N$) & \\
KEK-E307 \protect\cite{Sato05} & & & $0.828\pm 0.056\pm 0.066$ & &
\end{tabular}
}
\end{center}
\end{table}

%\begin{table}
%\begin{center}
%\caption{Non-mesonic weak decay rates (in units of the free $\Lambda$ decay width)
%and asymmetry parameters predicted for $^{11}_\Lambda$B}
%\label{boro}
%\begin{tabular}{c c c c c c}
%\mc {1}{c}{} &
%\mc {1}{c}{$\Gamma_n$} &
%\mc {1}{c}{$\Gamma_p$} &
%\mc {1}{c}{$\Gamma_{NM}$} &
%\mc {1}{c}{$\Gamma_n/\Gamma_p$} &
%\mc {1}{c}{$a_\Lambda$} \\ \hline
%$\pi$        & 0.067 & 0.619 & 0.685 & 0.108 & $-0.353$\\
%$\pi+K$      & 0.137 & 0.308 & 0.445 & 0.447 & $-0.773$\\
%$\pi+K+2\pi$ & 0.187 & 0.427 & 0.613 & 0.438 & $-0.074$\\
%OME          & 0.179 & 0.408 & 0.587 & 0.439 & $-0.809$\\
%OME$+2\pi$   & 0.201 & 0.425 & 0.627 & 0.474 & $-0.181$\\ \hline
%KEK-E508 \protect\cite{OutaVa} &  &  &  &  &
%\end{tabular}
%\end{center}
%\end{table}
%\begin{table}[h]
%\begin{center}
%\caption{Non-mesonic weak decay rates (in units of the free $\Lambda$ decay width)
%and asymmetry parameters predicted for $^{12}_\Lambda$C}
%\label{carbono}
%\begin{tabular}{c c c c c c}
%\mc {1}{c}{} &
%\mc {1}{c}{$\Gamma_n$} &
%\mc {1}{c}{$\Gamma_p$} &
%\mc {1}{c}{$\Gamma_{NM}$} &
%\mc {1}{c}{$\Gamma_n/\Gamma_p$} &
%\mc {1}{c}{$a_\Lambda$} \\ \hline
%$\pi$        & 0.066 & 0.751 & 0.817 & 0.088 & $-0.350$\\
%$\pi+K$      & 0.134 & 0.371 & 0.505 & 0.363 & $-0.643$\\
%$\pi+K+2\pi$ & 0.182 & 0.521 & 0.703 & 0.349 & $-0.093$\\
%OME          & 0.175 & 0.491 & 0.667 & 0.357 & $-0.698$\\
%OME$+2\pi$   & 0.194 & 0.529 & 0.722 & 0.266 & $-0.207$\\ \hline
%KEK-E508 \protect\cite{OutaVa} &  &  &  &  &
%\end{tabular}
%\end{center}
%\end{table}

The weak decay observables predicted for $^5_\Lambda$He, $^{11}_\Lambda$B
and $^{12}_\Lambda$C are compared with recent data obtained at KEK in
Table~\ref{nmdw}.
%In Table~\ref{nmdw}, our results for the decay widths and the intrinsic
%asymmetries are reported for $^5_\Lambda$He, $^{11}_\Lambda$B and
%$^{12}_\Lambda$C. Recent experimental data, obtained at KEK,
%are also given for comparison.
Concerning the ratio $\Gamma_n/\Gamma_p$, only experimental
results from nucleon-nucleon coincidence experiments
\cite{OutaVa,Ka06,Kim06} are quoted. These data should be
preferred over the ones obtained from single-nucleon studies; the
former are less affected by nucleon final state interactions and
two-nucleon-induced decays than the latter \cite{Bau06}. In
Table~\ref{nmdw} we also list theoretical determinations from
studies of nucleon final state interactions, with or without the
inclusion of the two-nucleon induced decay mode. They have been
obtained in Ref.~\cite{prl-prc} by fitting data on nucleon-nucleon
spectra from Refs.~\cite{OutaVa,Ka06,Kim06}. Note that the
determinations of $\Gamma_n/\Gamma_p$ by Ref.~\cite{prl-prc} are
sometimes significantly smaller than the corresponding
experimental results quoted in Table~\ref{nmdw}. This signals the
importance of final state interactions and two-nucleon induced
decays ---neglected \cite{Ka06} or accounted for in an approximate
way \cite{Kim06} in experimental analyses--- even when extracting
the ratio from nucleon--nucleon coincidence observables.

We start recalling the results of the OME model, including the
exchange of the mesons belonging to the ground state pseudoscalar
and vector octets. These results slightly differ from those of
Ref.~\cite{Pa02} due to the use here of numerically improved
correlated $NN$ wave functions. For the three hypernuclei under
study, both the  neutron-to-proton ratio and the total non-mesonic
width are reasonably reproduced within the OME model, especially
if one considers that non-negligible two-nucleon induced decay
rates [$\Gamma_2/(\Gamma_n+\Gamma_p)\simeq 0.20$ for
$^5_\Lambda$He and $\Gamma_2/(\Gamma_n+\Gamma_p)\simeq 0.25$ for
$^{11}_\Lambda$B and $^{12}_\Lambda$C] \cite{prl-prc,Al00PRC}
should be taken into account as well.  The values predicted
for the intrinsic asymmetries are large and negative, whereas small results,
compatible with zero, have been reported by recent experiments for all three
hypernuclei.

The effects of uncorrelated ($2\pi$) and correlated
($2\pi/\sigma$) two-pion contributions are better visualized by
including them, sequentially, to those of the lighter mesons
($\pi$ and $K$). As it is well know, the dominant tensor component
in the one-pion-exchange mechanism disfavors neutron-stimulated
decays and produces very small $\Gamma_n/\Gamma_p$ values. The
addition of kaon-exchange reduces $\Gamma_{\rm NM}$ by about 40\%
while increasing $\Gamma_n/\Gamma_p$ to values compatible with
data. This result is also well-known, being mainly due to i) the
enhancement of the parity-violating $\Lambda N(^3S_1)\to
nN(^3P_1)$ transition contributing especially to neutron-induced
decays and ii) the reduction of the tensor component, which for
kaon-exchange has  opposite sign of the one for pion-exchange. The
size of the asymmetry is doubled and practically reaches the large
value of the OME model. In fact, for all observables, the pion-
plus kaon-exchange contributions already constitute a large
fraction of the OME result.

As expected from the size of their respective potentials, see
Fig.~14 of Ref.~\cite{Os01}, the uncorrelated two-pion-exchange
mechanism has a much larger influence than the correlated one. We
observe that the $2\pi$ contribution increases $\Gamma_p$
substantially and $\Gamma_n$ more moderately, hence giving rise to
a decrease of the $\Gamma_n/\Gamma_p$ ratio with respect to the
$\pi+K$ result, which is especially sizable in the case of $^5_\Lambda$He.
The $2\pi/\sigma$ contribution affects the partial rates mildly,
reducing $\Gamma_n$ by less than 10\% and $\Gamma_p$ by slightly
more than 10\%.  As it could be reasonably expected on the basis
of the masses of the mesons included in the adopted weak
transition potential, the exchange of two pions, both uncorrelated
and correlated, turns out to be the most relevant mechanism beyond
pion- and kaon-exchange, giving an appreciable contribution to the
non-mesonic rates.
%despite maintaining the ratio $\Gamma_n/\Gamma_p$ rather unaffected.
%Here, we anticipate that two-pion-exchange is of extreme
%importance for a proper evaluation of the decay asymmetries.

The effect of the two-pion exchange contribution is
larger than that found in \cite{Os01}, where this potential was built and
applied to the decay of hypernuclei within a local density approximation
approach. This is  probably due to a different implementation of short range
correlations. The use of a phenomenological $NN$ correlation function of the
type $1-j(q_cr)$, with  a cut-off momentum of $q_c=780$ MeV, reduces the
rates considerably \cite{Pa02}. We have checked that, in this situation, the
relative changes induced by the two-pion scalar-isoscalar contributions amount
to about half of those seen in the results of Table~\ref{nmdw}, which are
obtained using realistic $NN$ wave-functions.

The most spectacular change induced by the uncorrelated two-pion
mechanism is seen in the asymmetry parameter, which turns from
being large and negative to being small and positive.
Incorporating the $2\pi/\sigma$ mechanism brings some additional
changes, basically tempering out the above mentioned effects. The
remaining heavier mesons  produce very moderate changes in the
decay widths, while the asymmetry parameter, being built from
interferences, shows a much stronger sensitivity. Roughly
speaking, the incorporation of the scalar-isoscalar terms to the
OME model leaves the rates basically unaltered, while reducing
substantially the absolute value of the  intrinsic asymmetry in
such a way that the predictions for all weak decay observables are
in excellent agreement with the measured values.

We note that a proper comparison with the  observed asymmetries
requires to account for the final state interactions of the weak
decay nucleons as they go out of the residual nucleus, as done in
Ref.~\cite{asyth05}. However, before commenting on these effects
below, we analyze the changes on the asymmetry in terms of the
modifications induced by our isoscalar-scalar mechanism in the
various transition amplitudes.

 By applying appropriate projection operators to the elementary
$\Lambda N \to nN$ potential, it is possible to select, from the
{\it hypernuclear} transition amplitude, the contributions coming
from specific spin-space transitions, $^{2S+1} L_J \to
^{2S^\prime+1}\!\!L^\prime_J$.  For $^5_\Lambda$He, 
the resulting amplitudes are
denoted by the capital letters $A,B,C,D,E,F$ in complete analogy
with the notation $a,b,c,d,e,f$ used for the same amplitudes in
the two-body case. In order to disentangle the contributions to
the asymmetry coming from the various interferences, we also
perform calculations for specific pairs of transitions,  which
will be denoted as $AE$, $BC$, $BD$, $CF$ and $DF$. Table
\ref{tab:amplitudes} shows the size of each proton-induced decay
amplitude, including its sign, for
the OME and ${\rm OME}+2\pi+2\pi/\sigma$ models. The sum of the
modulus squared of these amplitudes builds up the corresponding
value of $\Gamma_p$. We also show the contribution of all possible
interferences between pairs of amplitudes to the asymmetry. The
sum of all these interferences produces the final result for the
intrinsic asymmetry.

\begin{table}
\begin{center}
\caption{Hypernuclear amplitudes and interference terms in the 
proton-induced decay of $^5_\Lambda$He.}
\label{tab:amplitudes}
\begin{tabular}{c c c c c}
 & \mc{1}{c}{~~Parity~~}& {~~Isospin~~} &
\mc {1}{c}{~~${\rm OME}$~~} & \mc {1}{c}{~~${\rm
OME}+2\pi+2\pi/\sigma$~~}
\\ \hline
$A:~^1S_0 \to ^1\!\!S_0$ & PC & 1 & $- 0.1044$ & $+0.0835$  \\
$B:~^1S_0 \to ^3\!\!P_0$ & PV & 1 & $+0.0057$ & $+0.0057$\\
$C:~^3S_1 \to ^3\!\!S_1$ & PC & 0 & $- 0.1399$ & $+0.1480$ \\
$D:~^3S_1 \to ^3\!\!D_1$ & PC & 0 & $- 0.1814$ & $- 0.1814$ \\
$E:~^3S_1 \to ^1\!\!P_1$ & PV & 0 & $+0.3833$ & $+0.3833$ \\
$F:~^3S_1 \to ^3\!\!P_1$ & PV & 1 & $+0.2234$ & $+0.2234$ \\
\hline
$\Gamma_p=\displaystyle\sum_{\alpha={A \dots F} } |\alpha|^2$ & & & $0.257$ & $0.275$ \\
\hline
$AE$ & & & $-0.2854$ & $+0.2112$  \\
$BC$ & & & $+0.0027$ & $-0.0033$  \\
$BD$ & & & $-0.0029$ & $-0.0027$ \\
$CF$ & & & $-0.0856$ & $+0.0405$ \\
$DF$ & & & $-0.2186$ & $-0.2046$ \\
\hline
$a_\Lambda$ & & & $-0.590$ & $+0.041$
\end{tabular}
\end{center}
\end{table}

Except for the negligible $B(^1S_0 \to ^3\!\!P_0)$ amplitude, the
other ones turn out to be of relevance in the determination of the
proton decay asymmetry. In the case of the OME model, the parity--conserving
amplitudes ($A$, $C$ and $D$) are negative, and the
parity--violating ones ($B$, $E$ and $F$) positive. We note that
the larger contributions to the asymmetry turn out to be negative
and correspond to the interferences between the $A$ and $E$
($AE$), the $D$ and $F$ ($DF$), and the $C$ and $F$ ($CF$)
amplitudes. Our two-pion scalar-isoscalar mechanism affects the
parity conserving amplitudes which are diagonal in $S$ and $L$,
namely $A$ and $C$. As we see, they even change their sign which, in
turn, transform the negative interferences $AE$ and $CF$ into
positive contributions that largely cancel the negative $DF$
interference. We note that the small reduction in magnitude of the
$BD$ and $DF$ contributions to the asymmetry is just a reflection
of the slight increase of the $\Gamma_p$ rate. As a consequence of
the above mentioned change of sign, 
the asymmetry of $^5_\Lambda$He turns from
being large and negative in the OME model to being slightly
positive in the ${\rm OME}$ plus chiral $2\pi+2\pi/\sigma$ model,
in perfect agreement with the experimental observations.

Complementing the one-pion exchange mechanism in the weak decay of
hypernuclei with that of two correlated pions, in the scalar
($2\pi/\sigma$)  and vector ($2\pi/\rho$) sectors, was considered
for the first time by Itonaga et al. \cite{Itproc}. The model was
later extended to incorporate the exchange of the $\omega$
\cite{It02} and $K$ \cite{It04} mesons. We should note that the
approximation scheme employed in these works is purely
phenomenological. Their mechanism is such that, at the weak
vertex, two pions are emitted  via an intermediate $N$ or $\Sigma$
baryon. These two pions couple to a $\sigma$ meson, which is
absorbed by a nucleon at the $\sigma NN$ strong vertex. The needed
$\sigma$ mass, $m_\sigma$, and $\sigma NN$ coupling, $g_{\sigma
NN}$, are taken from phenomenological fits of the strong $NN$
interaction, while the remaining unknown $\pi \pi \sigma$ coupling
is fitted to reproduce the decay rates of $p$-shell hypernuclei.
It is not clear whether this model includes in an effective
way the uncorrelated two-pion mechanism. Another phenomenological
and even simpler approach to $\sigma$-meson exchange is that
adopted in Refs.~\cite{Sa05,Ba06}, where the strong ${\sigma NN}$
coupling constant is taken equal to that of pion, $g_{\sigma NN} =
g_{\pi NN} \sim 13.2$, and the weak $\sigma \Lambda N$ vertex is
parametrized in terms of a parity-conserving ($A_{\sigma}$) and
parity-violating ($B_{\sigma}$) coupling constants that are
adjusted to reproduce some weak decay observables. In
Ref.~\cite{Sa05}, this $\sigma$ is added to one-pion-exchange,
one-kaon-exchange and the direct quark transition
induced by an effective four-quark Hamiltonian. In
Ref.~\cite{Ba06}, the $\sigma$ is added to a meson-exchange model
which is similar to the one considered here.

In contrast, the two-pion scalar-isoscalar contributions
considered in the present work are theoretically well grounded in
the sense that all the coupling constants are determined from
chiral meson-meson and meson-baryon Lagrangians and by imposing SU(3)
symmetry. The regularizing parameter is adjusted such that
$\pi\pi$ scattering data is reproduced from threshold to around a
center-of-mass energy of around 1 GeV, well beyond the $\sigma$ region.
Having such different origin, it becomes difficult to perform a
comparative analysis with the above phenomenolgical models. 
We will just point out some differences in the results.

The work of Ref.~\cite{It04} found that to reproduce small and
positive values of $a_\Lambda(^5_\Lambda{\rm He})$, as experiment
indicates, their $2\pi/\sigma$ potential \cite{Itproc} is too
strong and must be decreased to half of its calculated value,
hence producing an important reduction of the amplitude $A$. This
in turn reduces the negative $AE$ term and the asymmetry gets
dominated by their positive $F(C + D)$ term. It becomes clear that
the way of obtaining a positive asymmetry in Ref.~\cite{It04} is
radically different from what it is found in the present work.

The one-pion and one-kaon exchange potential of Ref.~\cite{Sa05} are
modified by a $\sigma$-exchange contribution whose
coupling constants are fitted to various observables in light nuclei.
It is found that small values of the parity-violating coupling
constant $(B_\sigma \sim 1)$ and a large value of the 
parity-conserving one $(A_\sigma \sim 4)$ would reproduce the measured
$\Gamma_{\rm NM}$ and $\Gamma_n/\Gamma_p$ in $^5_\Lambda$He, but the
asymmetry would turn positive and large, of the order of 0.6.
Another reasonable fit to the rates is found with $A_\sigma \sim
-1.5$, but then the asymmetry is very large and negative, close to
$-1$. This work concludes that the additional inclusion of the direct quark
 mechanism permits finding a solution that reproduces both the partial rates
 and the asymmetry, in which case the values $A_\sigma \sim 4$ and
 $B_\sigma \sim 6.6$ are found.

Qualitatively similar results are found in Ref.~\cite{Ba06} in the
sense that their $\sigma$-exchange potential added to the full OME
exchange model can fit $\Gamma_{\rm NM}$ and $\Gamma_n/\Gamma_p$
but not the asymmetry. However, their solutions are
intrinsically very different, since the PV strength of the
$\sigma$ meson is dominant in Ref.~\cite{Ba06} ($B_\sigma/A_\sigma
\sim 10$ to 20) while the PC and PV $\sigma$ amplitudes are
of comparable strength in Ref.~\cite{Sa05} 
($B_\sigma/A_\sigma \sim 1$).
%($\mid B_\sigma/A_\sigma \mid \sim 1$).

Finally, we present in Table~\ref{asyFSI} our results for the
asymmetry after incorporating the effects of final state
interactions (FSI) on the emitted nucleons. These results are then
directly comparable with the observed asymmetries. We
show predictions for three hypernuclei and for the OME and
OME+$2\pi+2\pi/\sigma$ models. The first line in each case gives the asymmetry
in the absence of FSI and without applying any cut on the
kinetic energy of the emitted protons. The following lines
incorporate the effect of FSI for different energy cuts, namely
$T_p^{\rm th}=0$, 30 and 50 MeV, to accommodate to the
experimental conditions. A cut of $T^{\rm th}_p\sim 30$ to 50 MeV is
applied to $^5_\Lambda$He data, while $T_p^{\rm th}\sim 30$ MeV
for $^{11}_\Lambda$B and $^{12}_\Lambda$C.
%In Ref.~\cite{Ma05} a cut of $T^{\rm th}_p=30$ 
%MeV was applied, while $T_p^{\rm th}=30$ to 50 MeV in
%Ref.~\cite{Ma06}. 
We observe that, as in our OME study of
Ref.~\cite{asyth05}, the incorporation of FSI reduces the
magnitude of the observable asymmetry with respect to the intrinsic
asymmetry and that, as the kinetic energy cut is increased, 
$a^{\rm M}_\Lambda$ tends to recover the value of $a_\Lambda$.
The results of Table~\ref{asyFSI} show that the OME
model cannot reproduce the measured values of the asymmetries, while
the additional incorporation of the $2\pi + 2\pi/\sigma$ mechanism
provides asymmetry results in complete agreement with the
data for all hypernuclei.

\begin{table}
\begin{center}
\caption{Intrinsic and observable decay asymmetries predicted for
$^5_\Lambda$He, $^{11}_\Lambda{\rm B}$ and $^{12}_\Lambda{\rm C}$.}
\label{asyFSI}
\resizebox*{\textwidth}{!}{
\begin{tabular}{l c c c}
\mc {1}{c}{} & \mc {1}{c}{~~~~~~~$^5_\Lambda{\rm He}$~~~~~~~} &
\mc {1}{c}{~~~~~~~$^{11}_\Lambda{\rm B}$~~~~~~~} & \mc
{1}{c}{~~~~~~~$^{12}_\Lambda{\rm C}$~~~~~~~} \\ \hline
OME, $T^{\rm th}_p=0$ MeV & $-0.590$  & $-0.809$  & $-0.698$   \\
~~~~ FSI, $T^{\rm th}_p=0$ MeV    & $-0.260$  & $-0.173$  & $-0.145$   \\
~~~~ FSI, $T^{\rm th}_p=30$ MeV   & $-0.401$  & $-0.400$  & $-0.340$   \\
~~~~ FSI, $T^{\rm th}_p=50$ MeV   & $-0.455$  & $-0.554$  & $-0.468$   \\
%With FSI and $T^{\rm th}_p=70$ MeV   & $-0.493$  & $-0.689$  & $-0.564$  \\
\hline
${\rm OME}+2\pi+2\pi/\sigma$, $T^{\rm th}_p=0$ MeV & $+0.041$  & $-0.181$  & $-0.207$ \\
~~~~ FSI, $T^{\rm th}_p=0$ MeV    & $+0.021$  & $-0.038$  & $-0.048$   \\
~~~~ FSI, $T^{\rm th}_p=30$ MeV   & $+0.028$  & $-0.111$  & $-0.126$   \\
~~~~ FSI, $T^{\rm th}_p=50$ MeV   & $+0.030$  & $-0.173$  & $-0.179$   \\
%With FSI and $T^{\rm th}_p=70$ MeV   & $+0.032$  & $-0.251$  & $-0.228$  \\
\hline EXP \protect\cite{Ma05} & $0.11\pm 0.08\pm 0.04$ &
\mc {2}{c}{$-0.20\pm 0.26\pm 0.04$} \\
EXP \protect\cite{Ma06} & $0.07\pm 0.08^{+0.08}_{-0.00}$ 
& \mc {2}{c}{$-0.16\pm 0.28^{+0.18}_{-0.00}$}
\end{tabular}
}
\end{center}
\end{table}

%%%%%%%%%%%%%%%%%%%%%%%%%%%%%%%%%%
\section{Conclusion}
%%%%%%%%%%%%%%%%%%%%%%%%%%%%%%%%
\label{conclusion}

We have studied the non-mesonic weak decay of hypernuclei within a
one-meson-exchange model supplemented with the contributions of
the uncorrelated ($2\pi$) and correlated ($2\pi/\sigma$)
two-pion-exchange mechanisms. These last mechanisms are based on a
chiral unitary model which describes $\pi\pi$
scattering data up to around 1 GeV and are taken from Ref.~\cite{Os01}.
Our finite nucleus approach includes realistic strong
correlations both in the initial and final states and considers
the final state collisions of the nucleons in their way out of the
residual nucleus.

We have found that the two-pion-exchange mechanisms modify moderately the
partial decay rates but have a tremendous influence on the
asymmetry parameter, due to the change of sign of some relevant
amplitudes. The one-meson-exchange plus
two-pion-exchange model turns out to be able to reproduce satisfactory, not
only the total and partial hypernuclear weak decay rates, but also
the asymmetries observed in the angular distribution of 
protons emitted by polarized hypernuclei.

Recent studies on the validity of the $\Delta
I=1/2$ isospin rule in the non-mesonic decay
\cite{Sa05,Pa98,Al00,Gi01} have been of large interest, especially
due to their connections with the determination of
$\Gamma_n/\Gamma_p$ and the asymmetry parameter. Although this
kind of studies should be warmly supported, here we have to note
that, according to our results, based on pure $\Delta I=1/2$
$\Lambda N\to nN$ transitions, there appears to be no need for the
introduction of $\Delta I=3/2$ contributions to explain the
observed non-mesonic decay rates and asymmetries.

%%%%%%%%%%%%%%%%%%%%%%%%%%%%%
\section*{Acknowledgments}
%%%%%%%%%%%%%%%%%%%%%%%%%%%%%%%
This work is partly supported by the EU contract 
FLAVIAnet MRTN-CT-2006-035482, 
by the contract FIS2005-03142 from MEC (Spain) and
FEDER, by the INFN-MEC collaboration agreement number 06-36, and 
by the Generalitat de
Catalunya contract 2005SGR-00343. This research is part of the EU
Integrated Infrastructure Initiative Hadron Physics Project under
contract number RII3-CT-2004-506078. CC acknowledges support from
the fellowship BES-2003-2147 (MEC, Spain).

%%%%%%%%%%%%%%%%%%%%%%%%%%%%%%%


\begin{thebibliography}{100}
\bibitem{foot1}
%Here and in the following,
When comparing theoretical values of $a_\Lambda$ with data
for $a^{\rm M}_\Lambda$ we assume that, as obtained in Ref.~\cite{asyth05},
$a^{\rm M}_\Lambda/a_\Lambda\simeq 0.7$
for $^{5}_\Lambda$He and $\simeq 0.6$ for $^{11}_\Lambda$B and $^{12}_\Lambda$C.
\end{thebibliography}
\end{document}